\documentclass[12pt]{iopart}

\usepackage{iopams}
\usepackage{bm}
\usepackage{epsfig}
\usepackage{psfrag}

\newcommand{\bd}{\bm}

\begin{document}

\title[Functional renormalization group in the broken symmetry phase: momentum \dots]
{Functional renormalization group in the broken symmetry phase:
momentum dependence and two-parameter scaling  
of the  self-energy
}

\author{Andreas Sinner, Nils Hasselmann, and Peter Kopietz}
  
\address{Institut f\"{u}r Theoretische Physik, Universit\"{a}t
  Frankfurt,  Max-von-Laue Strasse 1, 60438 Frankfurt, Germany}

\date{September 12, 2007}

\begin{abstract}
We include    spontaneous symmetry breaking
into the functional renormalization group  equations
for the irreducible vertices of   Ginzburg-Landau theories 
by augmenting these equations by a flow equation for the order parameter, which is
determined from the requirement
that at each renormalization group
(RG) step the vertex with one external leg vanishes identically.
Using this strategy, we propose a simple truncation 
of  the coupled RG
flow equations for the vertices in
 the broken symmetry phase of the Ising universality class
in $D$ dimensions. Our truncation  yields
the full momentum dependence of the self-energy
$\Sigma ( \bd{k} )$ and interpolates between lowest order
perturbation theory at large momenta $  \bd{k} $ and the critical 
scaling regime for small $\bd{k}$.
Close to the critical point,
our method yields  the self-energy in the scaling form
 $\Sigma ( \bd{k} ) = k_c^2 \sigma^{-} ( | \bd{k} | \xi , | \bd{k} | / k_c)$, where $\xi$ is the
order parameter correlation length, $k_c$ is the  Ginzburg scale, and
$\sigma^{-} ( x , y )$ is a dimensionless two-parameter scaling function
for the broken symmetry phase which we
explicitly  calculate within our truncation.

\end{abstract}


\maketitle

\section{Introduction}
\label{sec:INTRO}

The functional renormalization group (FRG)
has been  invented  by Wegner and Houghton \cite{Wegner73}
in the early days of the renormalization group (RG) as a
mathematically 
exact formulation of the Wilsonian RG.
In the past decade this method has gained new attention.
While there are several equivalent formulations of the FRG
involving different types of generating functionals,
in many cases it is advantageous to
formulate the FRG in terms of the generating functional
$\Gamma$ of the one-particle irreducible vertices, which
can be obtained from the generating functional of the
connected Green functions via a Legendre transformation \cite{Wetterich93,Morris94}.
Two different  strategies of solving the  formally exact  FRG equation
for  the functional $\Gamma$ have been developed:
the first is based
on the combination of the local potential approximation (LPA) with the derivative
expansion \cite{Bagnuls02,Berges02}.
 This approach has been very successful to obtain accurate
results for critical exponents \cite{Bagnuls02,Berges02,Delamotte04,Pawlowski05} and is
convenient to describe the broken symmetry phase \cite{Diehl07},
because it is based on an expansion
in terms of invariant densities, which automatically
fulfill all symmetry requirements.

The other strategy, which was pioneered by Morris \cite{Morris94} and 
has been preferentially used in the condensed matter community 
to study non-relativistic
fermions~\cite{Kopietz01,Salmhofer01}, 
is based on the expansion of $\Gamma$  in powers of the fields,
leading to an infinite hierarchy of coupled integro-differential equations
for the one-particle irreducible vertices.
This approach has the advantage 
of providing information on the momentum- and frequency dependence
of the vertices. However, there have been only two conceptually different 
attempts to extend the hierarchy of FRG flow equations for the vertices
arising from the field expansion into the broken symmetry phase.
One possibility is  to include  
a small symmetry-breaking component 
into the initial condition for the self-energy \cite{Salmhofer04,Honerkamp05,Gersch05}
and to check whether this component evolves to a 
macroscopic value as the  RG flow is integrated.
A disadvantage of this scheme is that 
the order parameter field and its fluctuations do not explicitly appear
and that one even has to invest some effort
to recover simple mean-field results.

Another possibility to extend the FRG flow equations for the
irreducible vertices into the broken symmetry  phase
was proposed in \cite{Schuetz06} (see also \cite{Ellwanger94}).
The basic idea is
to augment the hierarchy of flow equations for the vertices
by an additional equation for the flowing  order parameter, 
which is obtained from the requirement that, at each stage of the 
RG flow, the vertex with one external leg
vanishes identically.
The purpose of this work is to show how this
strategy works in practice.
For simplicity we shall consider here a 
simple classical  scalar $\varphi^4$-theory, 
describing the critical behavior of the Ising universality class;
generalizations of our method to quantum  systems
are straightforward.  For example, this method has recently been used to 
study superconductivity in the 
attractive electron gas, where the flow equation
for the order parameter  is equivalent to
a generalized BCS gap equation 
including fluctuation corrections \cite{Lerch07}.

Apart from showing how symmetry breaking can
be taken into account in the field expansion,
we present in this work two additional new results:
on the one hand, we propose a simple truncation of the
exact hierarchy of the flow equations for the irreducible vertices
in the broken symmetry  phase 
which yields the full momentum dependence of the self-energy $\Sigma ( \bd{k} )$, interpolating between the perturbative regime
for large momenta $k \equiv | {\bd{k}}|$ 
and the critical regime for $k \rightarrow 0$.
Different
strategies to calculate the ${\bd{k}}$-dependence of the self-energy (and, 
more generally, the momentum dependence of the higher order vertices) 
has recently been developed in \cite{Ledowski04} and
in \cite{Blaizot06}. 
On the other hand, we present in this work an  approximate
calculation of the {\it{two-parameter}} scaling function 
$\sigma^{-} ( k \xi , k / k_c )$  describing the scaling of
the self-energy of the system slightly below the critical temperature $T_c$.  
Here $\xi$ is the order parameter correlation length, and $k_c$ is the
Ginzburg scale, which  remains finite at the critical point \cite{Amit74}.
The corresponding scaling function $\sigma^{+} ( k \xi , k / k_c )$ 
in the symmetric phase (i.e., for temperature $T > T_c$) 
has recently been discussed in \cite{Hasselmann07}.
The fact that 
the Ginzburg scale $k_c$ appears in the scaling of
thermodynamic variables 
has been discussed in several recent 
works~\cite{Anisimov95,Pelissetto02,Bagnuls02},
but apparently has been ignored 
in the older RG  literature \cite{Halperin68,Ma73,Fisher74}.
For models with weak interactions, a
universal regime, covering
the complete crossover from the
vicinity of the Gaussian fixed point to the vicinity of the
Wilson-Fisher-fixed point,
exists
which can be described completely within a two parameter 
scaling theory~\cite{Anisimov95,Pelissetto02,Bagnuls02}. 
For the weakly interacting Bose gas at criticality the 
one-parameter scaling function $\sigma_{\ast} ( k / k_c ) =
\sigma^{-} ( \infty , k / k_c )$ has been calculated in 
\cite{Baym99,Zinnjustin00,Ledowski04,Blaizot06}.
However, the
full implications of a finite scale  $k_c$ at criticality and the resulting
{\it{two-parameter}} scaling theory of the correlation function
away from criticality  has only recently been examined \cite{Hasselmann07}.
The field expansion of the FRG allows us to study the
complete momentum dependence of correlation functions 
and is thus ideally suited to explore
the extended universality near the
critical temperature $T_c$.

The structure of the paper is as follows. 
In Sec.~\ref{sec:ERG} we formulate the exact FRG flow equations for the running 
order parameter $M_\Lambda$ and the momentum dependent self-energy
$\Sigma_\Lambda(\bd{k})$. 
Guided  by the LPA,
in Sec.~\ref{sec:TRUNC} a truncation scheme 
of the hierarchy of flow equations for the irreducible vertices
in the broken symmetry phase is introduced,
which allows us to calculate the momentum dependent self-energy $\Sigma ( \bd{k} )$.
In Sec.~\ref{sec:2P} we derive the two-parameter scaling function for
the self-energy using  different approximations.
We end in Sec.~\ref{sec:CONC} with a brief summary and mention 
further applications of our method.

\section{Exact RG flow equations in the broken symmetry phase}
\label{sec:ERG}
Our starting point  is the following classical action,
 \begin{eqnarray}
 S [ \varphi ]  =
\int d^D{r} \left[
\frac{1}{2}({\boldsymbol \nabla} \varphi)^2
+\frac{r_{\Lambda_0}}{2}\varphi^2
+\frac{u_{\Lambda_0}}{4!}\varphi^4
\right]
,
 \label{eq:Sdef}
\end{eqnarray}
where an ultraviolet cutoff $\Lambda_0$ is implicitly understood.
In the broken symmetry phase the Fourier transform
$ \varphi_{\bd{k}}$ has a finite vacuum expectation value,
 \begin{equation}
  \varphi_{\bd{k}} = \varphi_{\bd{k}}^0 + \Delta \varphi_{\bd{k}} \; , \; 
\varphi_{\bd{k}}^0 = (2 \pi )^D \delta ( \bd{k}  ) M
 \; ,
 \end{equation}
with $ \langle  \Delta \varphi_{\bd{k}} \rangle =0$.
By substituting this expression into (\ref{eq:Sdef}) and expanding in 
powers of 
$\Delta \varphi_{\bd{k}}$, we generate also terms
involving one and three powers of 
$\Delta \varphi_{\bd{k}}$.
Demanding that the vertex $\Gamma^{(1)}_{\Lambda_0}$ associated with  the term 
linear in $\Delta \varphi_{\bd{k}}$ should vanish,
we obtain the magnetization in the Landau approximation,
 \begin{equation}
  M_{\Lambda_0} = \left\{ 
      \begin{array}{cc}
      0 & \mbox{for $ r_{\Lambda_0} > 0 $}  \\
      \sqrt{- 6 r_{\Lambda_0}/u_{\Lambda_0} } & \mbox{for $ r_{\Lambda_0} < 0 $}
    \end{array}
  \right.
  \; .
  \label{eq:M0landau}
\end{equation}
This will serve as the initial condition for the flow equation of the order
parameter in  our functional RG approach. 

To derive an  exact hierarchy of flow equations for the
vertices of our model we introduce a momentum cutoff $\Lambda$
into the free propagator separating fluctuations with small momenta $|\bd{k} | \lesssim \Lambda$
from those with large momenta $|\bd{k} | \gtrsim \Lambda$.
Differentiating the generating functional $\Gamma$
of the one-particle irreducible vertices
with respect to $\Lambda$ and expanding $\Gamma$ in powers of the
fields, we obtain a formally exact hierarchy of
FRG flow equations for the vertices \cite{Morris94}.
To take into account symmetry breaking, we follow the approach
proposed in \cite{Schuetz06} and demand that
for all values of $\Lambda$ the flowing vertex $\Gamma^{(1)}_{\Lambda}$ 
associated with
the term linear in the fluctuation $\Delta \varphi_{\bd{k}}$ vanishes.
This yields a renormalization group equation for the flowing order parameter
$M_{\Lambda}$.
We would like to calculate the true order parameter $M = \lim_{\Lambda \rightarrow 0}
M_{\Lambda}$ and the true momentum dependent single-particle 
Green function $G (\bd{k}) =  \lim_{\Lambda \rightarrow {0}} G_{\Lambda} ( {\bd{k}})$, 
which we parameterize in terms of an irreducible self-energy $\Sigma ( {\bd k} )$,
 \begin{equation}
G ( {\bd{k}} ) = \frac{1}{ {\bd{k}}^2 + \Sigma ( {\bd{k}} )}
 \; .
 \end{equation}
For our purpose it is convenient to include the
term proportional to $r_{\Lambda_0}$ in (\ref{eq:Sdef}) into the definition of 
the self-energy.
In the broken symmetry phase, the initial condition for the self-energy at 
scale $\Lambda = \Lambda_0$ is then
\begin{equation}
 \Sigma_{\Lambda_0} ( {\bd{k}} ) = r_{\Lambda_0}
 + \frac{u_{\Lambda_0} }{2} M_{\Lambda_0}^2 = \frac{u_{\Lambda_0}}{3} 
 M_{\Lambda_0}^2 = - 2 r_{\Lambda_0}
 \; ,
 \end{equation}
where $M_{\Lambda_0}$ is given in
(\ref{eq:M0landau}).
As we reduce the cutoff, the evolution of the self-energy
is determined by the following exact
RG flow equation \cite{Schuetz06,Schuetz05},
\begin{eqnarray}
\fl
  \partial_{\Lambda} \Sigma_{\Lambda} ( \bd{k} )
   =  - \frac{1}{2} 
\int \frac{d^D k^\prime}{(2\pi)^D}
  \dot{G}_{\Lambda} ( \bd{k}^{\prime} ) \Gamma^{(4)}_{\Lambda} (
  \bd{k}^{\prime} , - \bd{k}^{\prime} , \bd{k} , - \bd{k} )
  \nonumber 
  \\
    -   \int \frac{d^D k^\prime}{(2\pi)^D}
  \dot{G}_{\Lambda} ( \bd{k}^{\prime} ) 
  {G}_{\Lambda} ( \bd{k}^{\prime} + \bd{k} ) \Gamma^{(3)}_{\Lambda} (
  \bd{k} , - \bd{k} - \bd{k}^{\prime} , \bd{k}^{\prime} )
  \Gamma^{(3)}_{\Lambda} (
  - \bd{k}^{\prime} , \bd{k} + \bd{k}^{\prime} , - \bd{k} )
  \nonumber 
  \\ 
+
    ( \partial_{\Lambda} M_{\Lambda} ) \Gamma_{\Lambda}^{(3)} ( \bd{k} , - \bd{k} , 0 )
  \; , 
  \label{eq:flowGamma}
\end{eqnarray}
while the flowing order parameter $M_{\Lambda}$ satisfies \cite{Schuetz06}
\begin{eqnarray}
  ( \partial_{\Lambda} M_{\Lambda} ) \Sigma_{\Lambda} (0)
  & = & -\frac{1}{2} \int \frac{d^D k}{(2\pi)^D}
\dot{G}_{\Lambda} ( \bd{k} )
  \Gamma^{(3)}_{\Lambda} ( \bd{k} , - \bd{k} , 0 )
 \; .
  \label{eq:orderflow}
\end{eqnarray}
The flow equations for the three-point vertex 
$\Gamma^{(3)}_{\Lambda}  ( \bd{k}_1 , \bd{k}_2 , \bd{k}_3)$ and 
the four point vertex 
$\Gamma^{(4)}_{\Lambda}    ( \bd{k}_1 , \bd{k}_2 , \bd{k}_3, \bd{k}_4 )   $
in the presence of symmetry breaking
have been written down diagrammatically in  
\cite{Schuetz06}. In the present work we shall not need these equations.

In the broken symmetry phase the proper choice of the cutoff procedure
is a delicate matter.
The simplest choice is perhaps a sharp cutoff in momentum 
space, where the
propagator is for $ | {\bd{k}} | < \Lambda_0 $ given by~\cite{Morris94}
\begin{equation}
  {G}_{\Lambda} ( \bd{k} ) = \frac{ \Theta ( | \bd{k} | -  \Lambda )}{ 
     \bd{k}^2 + \Sigma_{\Lambda} ( \bd{k} ) }
  \; ,
\end{equation}
and the corresponding single-scale propagator is
\begin{equation}
  \dot{G}_{\Lambda} ( \bd{k} ) =- \frac{ \delta ( | \bd{k} | - \Lambda )}{ 
\Lambda^2 +
    \Sigma_{\Lambda} ( \bd{k} ) }
  \; .
\end{equation}
While in the symmetric phase the sharp cutoff 
is very convenient \cite{Morris94,Ledowski04}, it leads to technical complications 
in the broken symmetry phase
(see the discussion after (\ref{eq:etadotGamma}) below).
These can be avoided if we use
a smooth cutoff procedure, which we implement 
via an additive regulator
$R_{\Lambda} ( {\bd{k}} )$ in the inverse
propagator \cite{Berges02}. The cutoff dependent propagator
is then
\begin{equation}
  {G}_{\Lambda} ( \bd{k} ) = \frac{1}{ 
     \bd{k}^2  
 + \Sigma_{\Lambda} ( \bd{k} ) +R_{\Lambda} ( {\bd{k}} )      }
  \; ,
 \label{eq:GRdef}
\end{equation}
and corresponding single-scale propagator is
 \begin{equation}
  \dot{G}_{\Lambda} ( \bd{k} ) = [-  \partial_{\Lambda}
 R_{\Lambda} ( \bd{k} ) ] G^2_{\Lambda} ( \bd{k} )
 \; .
 \end{equation}
At this point it is not necessary to completely specify the 
cutoff function $R_{\Lambda} ( \bd{k} )$, 
except that we
require it to be of the form \cite{footnote1,footnote2}
 \begin{equation}
 R_{\Lambda} ( {\bd{k}} ) = ( 1 - \delta_{ \bd{k}, 0}) 
 \Lambda^2 Z_l^{-1} R ( {\bd{k}}^2 / \Lambda^2  )
 \; ,
 \label{eq:Rdef}
\end{equation}
where  $R ( x )$ is some dimensionless function
satisfying $R ( \infty ) = 0$ and $R (0 )=1$.
The inverse of the flowing wave-function renormalization factor is given by
 \begin{equation}
 Z_l^{-1} =  1 + 
 \left. \frac{ \partial \Sigma_{\Lambda} ( \bd{k} ) }{  
\partial  \bd{k}^2  } 
 \right|_{\bd{k}^2 =0} 
 \; ,
 \end{equation}
where $l = - \ln ( \Lambda  / \Lambda_0 )$.
The introduction of $Z^{-1}_l$ in (\ref{eq:Rdef})
is necessary to preserve the re-parametrization invariance
of physical quantities under a rescaling of the fields \cite{Berges02,footnote2}.
For explicit calculations we shall use the
Litim cutoff \cite{Litim01},
 \begin{equation}
 R (x ) = (1-x ) \Theta ( 1 - x )
 \; .
 \label{eq:RLitim}
 \end{equation}
Another popular choice is \cite{Berges02}
 \begin{equation}
 R ( x ) = \frac{ x}{e^x -1 }
 \label{eq:RBerges}
 \; ,
 \end{equation}
which has the advantage of being analytic, but leads to more complicated integrals.

\section{Truncated  flow equation for the 
momentum dependent self-energy}
\label{sec:TRUNC}

The right-hand side of (\ref{eq:flowGamma})   depends on the vertices
$\Gamma^{(3)}_{\Lambda}$ and $\Gamma^{(4)}_{\Lambda}$
with three and four external legs, 
which satisfy more complicated flow equations \cite{Schuetz06,Schuetz05}
involving higher order vertices.
Keeping in line with the derivative expansion
for the effective action \cite{Berges02} we 
truncate the hierarchy as follows \cite{Schuetz06},
\numparts
\begin{eqnarray}
  \Sigma_{\Lambda} ( 0 ) & \approx & \frac{u_\Lambda}{3} M_{\Lambda}^2
  \label{eq:G2trunc}
  \; ,
  \\
  \Gamma^{(3)}_{\Lambda} ( \bd{k}_1  , \bd{k}_2 , \bd{k}_3 ) 
  & \approx & u_\Lambda M_{\Lambda}
  \label{eq:G3initialtrunc}
  \; ,
  \\
  \Gamma^{(4)}_\Lambda  ( \bd{k}_1  , \bd{k}_2 , \bd{k}_3 , \bd{k}_4 ) 
  & \approx & u_\Lambda 
  \label{eq:G4initialtrunc}
  \; .
\end{eqnarray}
\endnumparts
This truncation of the field expansion 
is motivated by the LPA with quartic approximation
for the effective potential  $U_{\rm eff}$,
where one approximates the generating
functional $\Gamma$ by \cite{Berges02}
\begin{equation}
\Gamma [\varphi] \approx 
\int d^D r U_{\rm eff}[\varphi^2(\bd{r})], 
\label{eq:EffInt}
\end{equation}
with 
\begin{equation}
U_{\rm eff} [\varphi ]\approx \frac{u_\Lambda}{4!}
\big[\varphi^2-M^2_\Lambda\big]^2.\label{eq:LPA}
\end{equation}
The completely local character of all correlations in the
LPA is known to be a surprisingly good approximation
in the scaling regime close to criticality \cite{Berges02}. Outside
the critical regime the LPA fares less well and in general
cannot reproduce the structure known from perturbation theory (in the
case considered here, only the leading order from perturbation theory
will be reproduced).
The LPA combined with a derivative expansion converges best if one 
expands around the local minimum of $U_{\rm eff}$, see \cite{Aoki98}. 
The condition that $M_\Lambda$ is the flowing minimum leads to a flow 
equation for $M_\Lambda$, which in the field expansion
leads to (\ref{eq:G2trunc}--\ref{eq:G4initialtrunc}).
In this truncation, the exact flow equation (\ref{eq:orderflow})
for the order parameter reduces to
\begin{equation}
  \partial_{\Lambda} M_\Lambda^2 = - 3\int \frac{d^D k}{(2\pi)^D}
\dot{G}_{\Lambda} ( \bd{k} ),
  \label{eq:flowM}
\end{equation}
while our flow equation (\ref{eq:flowGamma}) for the self-energy
becomes
\begin{eqnarray}
  \hspace{-1.5cm}
\partial_{\Lambda} \Sigma_{\Lambda} ( \bd{k} )
  & = & \frac{u_{\Lambda}}{2}\int \frac{d^D k^\prime}{(2\pi)^D}
  \dot{G}_{\Lambda} ( \bd{k}^{\prime} ) 
  +\frac{u_\Lambda}{2}
  \partial_{\Lambda} M^2_{\Lambda}
   -   u_{\Lambda}^2 M_{\Lambda}^2 \int \frac{d^D k^\prime}{(2\pi)^D}
  \dot{G}_{\Lambda} ( \bd{k}^{\prime} ) 
  {G}_{\Lambda} ( \bd{k}^{\prime} + \bd{k} ) 
 \nonumber
 \\
 & = & - u_{\Lambda}  \int \frac{d^D k^\prime}{(2\pi)^D}
  \dot{G}_{\Lambda} ( \bd{k}^{\prime} ) 
 -u_{\Lambda}^2 M_{\Lambda}^2 \int \frac{d^D k^\prime}{(2\pi)^D}
  \dot{G}_{\Lambda} ( \bd{k}^{\prime} ) 
  {G}_{\Lambda} ( \bd{k}^{\prime} + \bd{k} ) 
  \; .
  \label{eq:flowGamma2}
\end{eqnarray}
By demanding that the flow of $\Sigma_{\Lambda} ( 0 )$ is consistent
with our truncation (\ref{eq:G2trunc}) we obtain the flow equation
for the effective interaction,
\begin{equation}
  \partial_{\Lambda} u_{\Lambda} = - 3 u_{\Lambda}^2 \int \frac{d^D k}{(2\pi)^D}
  \dot{G}_{\Lambda} ( \bd{k} ) G_{\Lambda} ( \bd{k} )
  \; .
  \label{eq:flowu}
\end{equation}
The above equations  (\ref{eq:flowM}--\ref{eq:flowu})
form a closed system of coupled integro-differential equations 
for the order parameter $M_{\Lambda}$, the effective interaction $u_{\Lambda}$, and 
the momentum dependent self-energy
$\Sigma_{\Lambda} ( {\bd{k}} )$.
In contrast to the LPA, these equations can be used to calculate
the full ${\bd{k}}$-dependence of $\Sigma({\bd k})$. 
Our truncation is similar in spirit but not identical to the
more elaborate truncation proposed in \cite{Blaizot06}, who also 
used the LPA as a guide to propose 
a truncation of the hierarchy of flow equations  for the momentum dependent
vertices generated in the field expansion. 
However, unlike  our
equation (\ref{eq:flowGamma}), 
the flow equation for the self-energy proposed
by Blaizot {\it{et al.}} \cite{Blaizot06} does not
involve  the flowing order parameter, 
because
these authors approach the critical point using an expansion around
the symmetric state.

At this point it is convenient to rescale all quantities to 
reveal their scaling dimensions. 
We define dimensionless momenta ${\bd{q}} =  {\bd{k}} / \Lambda$ and the dimensionless 
coupling constants
\begin{eqnarray}
  {u}_l & = &  K_D Z_l^2 \Lambda^{D-4} u_{\Lambda}
 \label{eq:tildeudef}
 \; ,
 \\
{M}_l^2 & = & \frac{ M^2_{\Lambda} }{Z_l K_D \Lambda^{D-2}} 
 \label{eq:tildeMdef}
 \; ,
 \end{eqnarray}
which are considered to be functions of $l = - \ln ( \Lambda / \Lambda_0 )$.
Here $K_D$ is defined by
\begin{equation}
 K_D=\frac{\Omega_D}{(2\pi)^D}=\frac{2^{1-D}}{ \pi^{D/2} \Gamma(D/2)} \; ,
\end{equation}
where $\Omega_D$ is the surface area of the $D$-dimensional unit sphere.
We also define the rescaled exact propagator,
 \begin{eqnarray}
 {G}_l ( {\bd{q}} ) & = & \frac{ \Lambda^2}{ Z_l} G_{\Lambda} ( {\bd{k}} )
 \nonumber
 \\
 & = & \frac{1}{ Z_l  {\bd{q}}^2  + 
 {\Gamma}^{(2)}_l ( \bd{q} )   +  R ( {\bd{q}}^2 )   }
 \; ,
 \label{eq:Gscale}
 \end{eqnarray}
and the corresponding single-scale propagator
 \begin{equation}
 \dot{G}_l ( {\bd{q}} ) = \dot{R}_l ( \bd{q} ){G}_l^2 ( {\bd{q}} ) ,
 \end{equation}
where
 \begin{equation}\label{eq:RPoint}
  \dot{R}_l ( \bd{q}  ) = - \frac{ Z_l}{ \Lambda} \partial_{\Lambda} R_{\Lambda} (
 {\bd{k}} ) = - ( 2 - \eta_l)   R ( \bd{q}^2 )  + 2  \bd{q}^2 R^{\prime} ( \bd{q}^2 ) .
 \end{equation}
Here $R^{\prime} ( x ) = d R ( x ) / dx$ and
 $\eta_l = - \partial_l \ln Z_l$ is the flowing anomalous dimension.
For the Litim cutoff 
$R^{\prime} ( x )   = -  \Theta ( 1 - x )$, so that
 \begin{equation}
\dot{R}_l ( \bd{q}  ) =  [ - 2  + \eta_l (1 - \bd{q}^2) ]  \Theta ( 1 - \bd{q}^2 ).
\label{eq:RPoint2} 
\end{equation}
The rescaled propagator (\ref{eq:Gscale}) depends on
the rescaled irreducible self-energy,
 \begin{equation}
{\Gamma}_l^{(2)} ( \bd{q} ) = 
\frac{ Z_l}{ \Lambda^2} \Sigma^{}_{\Lambda} ( \bd{k} )
 \label{eq:tildeGamma2}
 \; .
 \end{equation}
By construction, the 
constant part of the rescaled self-energy is
 \begin{equation} \label{eq:rhodef}
  {\Gamma}^{(2)}_l ( 0 )  = \frac{{u}_l}{3} {M}_l^2 = \frac{Z_l}{ \Lambda^2}
  \frac{{u}_\Lambda}{3} {M}_\Lambda^2
 \; .
 \end{equation}
The flow of $ {\Gamma}^{(2)}_l ( 0 )  $ is thus determined
by the flow of ${M}_l^2$ and ${u}_l$, which in our truncation is given by
\begin{equation}
 \partial_l {M}_l^2 =  ( D-2 + \eta_l ) {M}_l^2 + 3 
 \int_{\bd{q}} \dot{G}_l ( \bd{q} )
 \label{eq:flowtildeM}
 \; ,
 \end{equation}
 \begin{equation}
 \partial_l {u}_l =  ( 4-D - 2 \eta_l ) {u}_l + 3 {u}_l^2 
 \int_{\bd{q}} \dot{G}_l ( \bd{q} ) {G}_l  ( \bd{q} )
 \label{eq:flowtildeu}
 \; ,
 \end{equation}
where $\int_{\bd q}=\Omega_D^{-1}\int
d^D{q}$.
To calculate $\eta_l$, we need the momentum dependent part of the rescaled self-energy,
 \begin{equation}
 {\gamma}_l ( {\bd{q}} ) = 
  {\Gamma}_l^{(2)} ( \bd{q} ) -
 {\Gamma}_l^{(2)} ( 0 ) 
 \; ,
 \end{equation}
which satisfies
 \begin{equation}
 \partial_l {\gamma}_l ( \bd{q} )   =  ( 2  -  \eta_l - \bd{q} \cdot \nabla_{\bd{q}} ) 
 {\gamma}_l ( \bd{q} )  +  \dot{\gamma}_l ( \bd{q} )
 \label{eq:flowtildesigma}
 \; ,
 \end{equation}
where
 \begin{equation}
 \dot{\gamma}_l ( \bd{q} ) = {u}_l^2 {M}_l^2 
 \int_{\bd{q}^{\prime} }  \dot{G}_l ( \bd{q}^{\prime} ) [{G}_l  ( \bd{q}^{\prime}
 + \bd{q} ) - {G}_l  ( \bd{q}^{\prime}  ) ] \; .
 \label{eq:Idef}
\end{equation}
The flowing anomalous dimension is then given by
 \begin{equation}
 \eta_l = \left. \frac{ \partial \dot{\gamma}_l ( {\bd{q}} ) }{\partial \bd{q}^2 } 
 \right|_{ \bd{q}^2 =0}  
 \; .
 \label{eq:etadotGamma}
 \end{equation}
The reason why using a sharp cutoff in the broken symmetry phase 
leads to technical complications
is that in this case
the expansion of the integral (\ref{eq:Idef}) for small
$ \bd{q} $ would start with a non-analytic term proportional
to $ | \bd{q} |$, which requires the introduction of an additional relevant coupling 
constant \cite{Ledowski04}.
On the other hand, 
with the Litim cutoff (\ref{eq:RLitim}) or the analytic cutoff (\ref{eq:RBerges})
the leading term in the expansion of 
$\dot{\gamma}_l ( \bd{q} )$ is proportional to $\bd{q}^2$.
In this case the expansion of
the right-hand side of (\ref{eq:Idef}) for small $\bd{q}$ yields
for the flowing anomalous dimension
 \begin{eqnarray}
\fl
 \eta_l =  - {u}_l^2 {M}_l^2 
 \int_{\bd{q} }  \dot{G}_l ( \bd{q}  ) 
 \Bigl\{ {G}_l^2 ( \bd{q} ) [ 1 +  R^{\prime} ( \bd{q}^{2} ) ]
+\frac{ \bd{q}^2}{D} \bigl[
 2 {G}_l^2 ( \bd{q} )  R^{\prime \prime} ( \bd{q}^{2} )-
 4  {G}_l^3 ( \bd{q} ) [ 1 +  R^{\prime} ( \bd{q}^{2} ) ]^2 \bigr]
 \Bigr\}
 \;,
 \nonumber
 \\
 & &
 \label{eq:floweta}
\end{eqnarray}
where $R^{\prime \prime} ( x ) = d^2 R ( x ) / d x^2$.
For the Litim cutoff $R^{\prime \prime} ( x ) = \delta (1-x )$.

\section{Two parameter scaling in the broken symmetry phase}
\label{sec:2P}
\subsection{Truncation with 
only marginal and relevant  couplings}
\begin{figure} \centering 
\includegraphics[height=5 cm, width=8 cm]{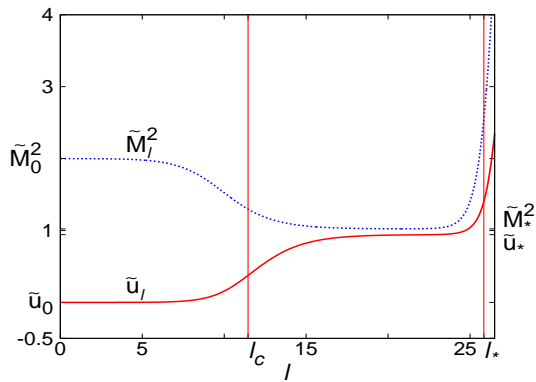}
\caption{  Typical non-critical flow of the coupling parameters 
$ M_l^2$ and $ u_l$ obtained from (\ref{eq:flowtildeMwithLitim}) 
and (\ref{eq:flowtildeuwithLitim}) in $D=3$.}
\label{fig:CouplingsFlow}
\end{figure}

\begin{figure} \centering 
\includegraphics[height=5 cm]{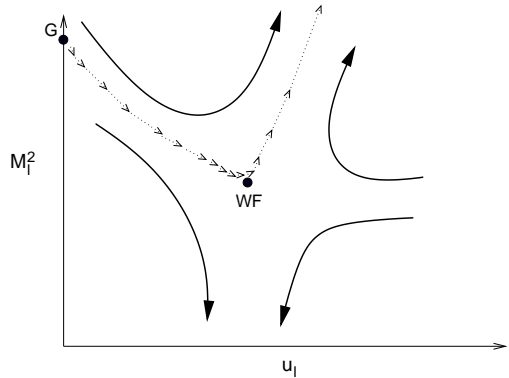}
\caption{Qualitative flow diagram for the couplings  $u_l$ and $ M^2_l$.
The arrows sketch a nearly critical flow, with the size of the arrows representing 
the velocity of the flow. The dots mark the Gaussian fixed point (G) and the
Wilson-Fisher fixed point (WF).}
\label{fig:FlowDiagram}
\end{figure}

In the simplest self-consistent approximation, we expand
the momentum dependent part ${\gamma}_l ( \bd{q} ) $ 
of the two-point vertex
on the right-hand side of our flow equations
(\ref{eq:flowtildeM}, \ref{eq:flowtildeu}, \ref{eq:flowtildesigma}) 
to first order in $\bd{q}^2$.
Since by definition
 \begin{equation}
 Z_l = 1 - \left. \frac{ \partial {\Gamma}_l^{(2)} ( {\bd{q}} ) }{\partial \bd{q}^2 } 
 \right|_{ \bd{q}^2 =0} = 1 - \left. \frac{ \partial {\gamma}_l ( {\bd{q}} ) }{\partial \bd{q}^2 } 
 \right|_{ \bd{q}^2 =0}
 \; ,
 \end{equation}
this amounts to approximating the propagator on the right-hand side of the flow
equations (\ref{eq:flowtildeM}, \ref{eq:flowtildeu}, \ref{eq:flowtildesigma})  
by
 \begin{equation}
 {G}_l ( \bd{q}  ) \approx  \frac{1}{ \bd{q}^2  +\rho_l  + 
  R ( \bd{q}^2 ) } ,
 \label{eq:Gtrunc}
 \end{equation}
where  
  \begin{equation}
\rho_l=\Gamma_l^{(2)}(0) = \frac{u_l}{3}  M_l^2,
 \end{equation}
see  (\ref{eq:rhodef}). 
The resulting system of flow equations for
the coupling constants ${M}_l^2$ and $ u_l$ together
the flow  equation $\partial_l Z_l = - \eta_l
Z_l$ for $Z_l$ are
equivalent to the quartic approximation for the effective 
potential with wave-function renormalization~\cite{Berges02}. 
Using  the Litim cutoff (\ref{eq:RLitim}),
equations (\ref{eq:flowtildeM}) and (\ref{eq:flowtildeu})  become
  \begin{eqnarray}
 \vspace{-.5cm}   \partial_l  {M}^2_l=(D-2+\eta_l) {M}^2_l-\frac{6(2+D-\eta_l)}{D(D+2)} G^2_l(0), &&
    \label{eq:flowtildeMwithLitim}
    \\
    \partial_l  u_l=(4-D-2\eta_l) u_l-\frac{6(2+D-\eta_l)}{D(D+2)} u^2_l  G^3_l(0), &&
    \label{eq:flowtildeuwithLitim}
  \end{eqnarray}
where $ G_l(0)\approx
\big[1+  \rho_l ]^{-1}$ is the rescaled propagator at zero momentum. 
Moreover, with the Litim cutoff the 
flowing anomalous dimension (\ref{eq:floweta}) is simply
\begin{equation}
\eta_l = \frac{1}{D} u_l^2  M^2_l  G^4_l(0).
\label{eq:flowetawithLitim}
\end{equation} 
Equations (\ref{eq:flowtildeMwithLitim}--\ref{eq:flowetawithLitim})
 form a closed system  of differential equations for $M_l^2$, $u_l$ and $\eta_l$
which can easily be solved  numerically. 
To find the flow along the critical surface, we need to fine tune 
carefully the initial values $u_0$ and $M_0^2$. 
A typical flow of the rescaled parameters as a function of $l$ is shown in 
figure \ref{fig:CouplingsFlow}, while in
figure \ref{fig:FlowDiagram} we show the flow schematically in the 
$\big( u_l, M_l^2\big)$-plane. 
The Wilson-Fisher fixed point in $D=3$ is in this approximation at 
$ u_*\approx 0.942$ and $ M^2_*\approx 1.022$.
As can be seen in  figure \ref{fig:FlowDiagram}, the couplings initially
flow very slowly and 
stay close to their initial values in the vicinity of 
the Gaussian fixed point.
At a characteristic scale $l_c$ they are rapidly attracted by
the Wilson-Fisher fixed
point where the flow is again almost stationary. Finally, 
at the scale $l_*$ all non-critical  RG trajectories
rapidly move away from the Wilson-Fisher point
and the $l$-dependence of the couplings $u_l$ and $M_l^2$ 
is determined by  their scaling dimensions,
$u_l \propto e^{ \epsilon l}$, $M_l^2 \propto e^{(D-2)l}$,
where $\epsilon=4-D$; the flow of the un-rescaled couplings $u_{\Lambda}$ and
$M^2_{\Lambda}$ then stops.

What determines the two characteristic scales $l_c$ and $l_*$? 
The momentum scale
$k_c=\Lambda_0 e^{-l_c}$ associated with $l_c$
measures the size of the Ginzburg critical 
region. For small initial values $u_0$ the logarithmic scale $l_c$ is 
given by \cite{Pelissetto02,Ledowski04,Amit74}
\begin{equation}
l_c\approx\frac{1}{\epsilon} \ln\Big(\frac{ u_*}{ u_0}\Big),
\label{eq:lc}
\end{equation}
where $ u_*$ is the value of $u_l$ at the Wilson-Fisher fixed point.  
This scale can be derived from (\ref{eq:flowtildeMwithLitim}) 
and (\ref{eq:flowtildeuwithLitim}) by approximating $ G_l(0) \approx 1$ and 
$\eta_l\approx 0$. 
In the intermediate regime $l_c \lesssim l \lesssim l_{\ast}$
the flowing  $M_l^2$ can then be replaced
by a constant $M_l^2 \approx M_*^2=6/[D(D-2)]$, while for all $l$
the solution of (\ref{eq:flowtildeuwithLitim}) can be approximated by 
\begin{equation}
\frac{ u_l}{ u_*} \approx \frac{1}{e^{\epsilon (l-l_c)}+1},
\end{equation}
where $ u_*=D \epsilon /6$.  The numerically obtained flow
shown in figure \ref{fig:CouplingsFlow}
further reveals that the scale $l_c$ is also characteristic for 
the $l$-dependence of $M_l^2$.   

Non-critical flows which describe the system at $T < T_c$ eventually obey  
$M^2_l/ M^2_*\gg  G^2_l(0)$ and $u_*/u_l \gg G^3_l(0)$. In that case,
the solution for $M_l^2$ and $u_l$ depend exponentially on $l$, 
corresponding to trivial scaling. The unrescaled variables $u_\Lambda$
and $M_\Lambda^2$ then approach finite limits and  also
the physical correlation length $\xi$, which is defined via
\begin{equation}
\xi^{-2}=\lim_{\Lambda\to 0} [Z_\Lambda \Sigma_\Lambda(0)],
\end{equation}
remains finite. 
The scale $l_*$  in figure {\ref{fig:CouplingsFlow}} is
related to $\xi$ via $\xi^{-1}=\Lambda_0 e^{-l_*}$ or equivalently
\begin{equation}
2 l_*=-\ln \big[\lim_{l\rightarrow\infty}e^{-2 l} \Gamma_l^{(2)}(0) \big].\label{eq:LStar}
\end{equation}

From the linearized flow around the Wilson-Fisher fixed point we obtain in 
$D=3$  the critical exponents $\nu \approx 0.553$ and $\eta \approx 0.099$. 
Very similar results are obtained using the analytic cutoff (\ref{eq:RBerges}). 
The poor comparison of our results to the established 
values $\nu =0.64$ and $\eta =0.044$ (see \cite{Berges02}) can be traced to 
the low order truncation of our effective potential (\ref{eq:LPA}), see
\cite{Aoki98, Canet03}. 
Close to $D=4$ we obtain in the leading order $\epsilon$-expansion  
$\eta \sim \epsilon^2/12$ and $\nu \sim 1/2+\epsilon/12$. 
While the result for $\nu$ is correct, the value for 
$\eta$ compares badly with the known expansion $\eta \sim \epsilon^2/54$.
This clearly shows the limitations  of a low-order effective potential approximation.

\subsection{FRG enhanced perturbation theory}
\label{sec:PFRG}

So far, we have truncated the self-energy retaining  only its marginal and relevant
parts. This is a good approximation for small momenta $k$. On the other hand,
for large $k$ this approximation cannot 
correctly reproduce the momentum dependence of the self-energy 
which arises from perturbation theory. 
To leading order in the relevant dimensionless  bare coupling
$\bar{u}_0  = u_{\Lambda_0} \xi^{4-D}$,
the perturbative  momentum dependence of $\Sigma(\bd{k})$
in the broken symmetry phase is
given by \cite{Hasselmann07}
\begin{equation}
\Sigma  (\bd{k})-\Sigma(0)=\xi^{-2}\Delta\sigma_0^{-} (k  \xi ),
 \label{eq:sigmaminuspert}
\end{equation}
with
\begin{eqnarray}
 \Delta\sigma_0^{-} (x)  =
  \frac{ 3\bar{u}_0}{2}  [ \chi ( 0 ) - \chi (x  ) ] + O ( \bar{u}_0^2 ) ,
 \label{eq:sigmaminus}
\end{eqnarray}
where 
 \begin{equation}
 \chi ( p ) =  \int  \frac{ d^D p^\prime}{ ( 2 \pi )^D}
  \frac{1}{ [\bd{p}^{\prime 2}  +1]  [(\bd{p}^{\prime } + \bd{p} )^2 +1 ]}.
 \label{eq:chipert}
 \end{equation}
(\ref{eq:sigmaminuspert}--\ref{eq:chipert}) are only 
accurate sufficiently far away from the critical
point where $\xi$ and the relevant dimensionless coupling
$\bar{u}_0  = u_{\Lambda_0} \xi^{4-D}$ are small.

We now present an improved approximation for the momentum 
dependent self-energy 
which we call  {\it{FRG enhanced perturbation theory}} since it 
embeds the perturbative expansion into a functional renormalization 
\cite{Wetterich96}
such
that it reproduces
exactly the leading order perturbative behavior  for large $\bd{k}$.
However, in contrast to perturbation theory,
FRG enhanced perturbation theory does not suffer from any  divergencies;
it yields an explicit description of 
the entire crossover to the critical regime and
gives reasonable results even at the critical  point.

\begin{figure} \centering 
\includegraphics[height=5 cm]{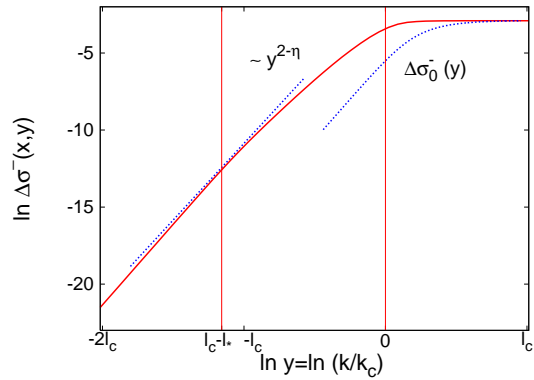}
\caption{   Typical behavior of the 
two-parameter scaling function $ \Delta \sigma^- ( x , y )$ defined 
in (\ref{eq:XYScFunction}) for $x = e^{l_{\ast} - l_c} y $.
The initial coupling parameters are $ u_0=0.005$ and $ M^2_0=1.9482092$, 
which yields $l_c\simeq 5.23$ and $l_*\simeq 11.29$.}
\label{fig:SigmaLog}
\end{figure}

Quite generally, the physical self-energy can be written 
as an integral over the entire RG trajectory
\cite{Hasselmann07},
\begin{equation}
\Sigma(\bd{k})-\Sigma(0)=\Lambda_0^2 \int_0^\infty dl e^{-2 l + \int_0^l d\tau\eta_\tau}\dot\gamma\big(e^l \bd{k}/\Lambda_0\big), 
\label{eq:sigma_general}
\end{equation}
where 
$\Sigma(0)=Z^{-1}\xi^{-2}$.
In general, the expression for
the inhomogeneity $\dot\gamma({\bd{q}})$ will depend also on the
momentum dependence of the three- and four-point irreducible vertices,
as can be inferred from (\ref{eq:flowGamma}).
To make progress, we employ the truncation 
(\ref{eq:G2trunc}-\ref{eq:G4initialtrunc}) which leads to the
approximation
(\ref{eq:Idef}) for the inhomogeneity $\dot\gamma({\bd{q}})$.
Only the momentum independent parts of the three- and four
point vertices enter and $\dot\gamma({\bd{q}})$ 
is then completely determined by the self-energy and the
order parameter alone. While this greatly simplifies the 
calculation of the self-energy since
it leads to a closed set of equations,
solving
(\ref{eq:sigma_general}) remains non-trivial since
the solution for the self-energy must be determined self-consistently. 
The FRG enhanced perturbation theory provides for
a non-self-consistent approximation to the solution of 
(\ref{eq:sigma_general}). In the FRG enhanced perturbation
theory,
the calculation of
the subtracted inhomogeneity $\dot\gamma({\bd{q}})$ via 
(\ref{eq:Idef}) is simplified by keeping only the first two
terms in a momentum expansion of the self-energy.
This amounts 
to the substitution
\begin{equation}
\Gamma_l^{(2)}(\bd{q})\rightarrow \Gamma_l^{(2)}(0) + (1-Z_l) \bd{q}^2
 \label{eq:substit} 
\end{equation}
and the approximation (\ref{eq:Gtrunc}) for the propagator
on the right-hand
side of  (\ref{eq:Idef}).
Within this 
approximation, the flow of $M_l^2$, $u_l$, and $\eta_l$ are determined
from
(\ref{eq:flowtildeMwithLitim}--\ref{eq:flowetawithLitim}).
A similar truncation strategy has been adopted 
in \cite{Ledowski04} for the symmetric phase of the $O(2)$-model, and in
\cite{Busche02} to calculate the spectral function of the
Tomonaga-Luttinger model. A comparison with the completely self-consistently
determined self-energy is presented at the end of this section where
we show that the error arising from the non-selfconsistency of the FRG 
solution is extremely small.
Perturbation theory is
recovered when the flow of the running couplings  is
approximated by their trivial $l$-dependence arising from their 
scaling dimensions.
The self-energy can now be expressed in terms of
a two-parameter scaling function,
\begin{equation}
\Sigma(\bd{k})=k_c^2 \sigma^- (x,y),
\end{equation}
with $x=k \xi$ and 
$y=k /k_c$. The ratio of these variables is then $x/y=e^{l_*-l_c}$.
If we introduce 
 \begin{equation}
\Delta\sigma^- (x,y) =\sigma^- (x,y)-  k_c^2  \Sigma(0)=
\sigma^- (x,y)-y^2/Zx^2,
 \end{equation} 
this leads to
\begin{eqnarray}
\fl \Delta\sigma^- (x,y) =
 \int_0^\infty dl e^{-2 (l-l_c)+\int_0^l d\tau \eta_\tau}\dot\gamma_l(e^{l-l_c}y)
= y^2\int_{ye^{-l_c}}^\infty dp \, p^{-3} Z^{-1}_{l_c+\ln (p/y)}
\dot\gamma_{l_c+\ln (p/y)}(p), 
\label{eq:XYScFunction}
\end{eqnarray} 
where we substituted $p=y e^{l-l_c}$ and used
$Z_l=e^{-\int_0^l d\tau \eta_\tau}$.
The asymptotic behavior of $\dot \gamma_l(\bd{q})$ 
for small $q$ is 
${\dot\gamma_l(\bd{q})}\approx\eta_l \bd{q}^2$.
For large $q$ it approaches a constant which, using the Litim cutoff, is
\begin{equation}
\lim_{q\to\infty}\dot\gamma_l(\bd{q})\approx
2 {u}^2_l M^2_l  G^3_l(0 )
\frac{(2+D-\eta_l)}{D(D+2)} \, .
\label{eq:UpAsymptGammaPoint}
\end{equation}
In $D=3$ the  function $\dot \gamma_l(\bd{q})$ 
can be calculated analytically for the Litim cutoff;
the result is given in the Appendix.
At the critical point $x\to\infty$  since the correlation length diverges,
so that the scaling function reduces to 
\begin{equation}
\Delta\sigma^- (\infty,y)=\sigma^- (\infty,y)=\sigma_*(y).
\end{equation}
The asymptotic behavior of the scaling function for both very small and very large $y$
follows directly from (\ref{eq:XYScFunction}).
For $y\ll 1$, i.e. in the critical long wavelength regime \cite{Hasselmann07},
 the lower limit of integration may be replaced by zero and all coupling parameters
may be replaced by their fixed point values.  Then we find 
\begin{equation}
\sigma_*(y)\approx A_D y^{2-\eta},
\end{equation}
where $\eta$ is the fixed point value of $\eta_l$ and
\begin{equation}
A_D=\int_0^\infty dp\; p^{\eta-3}\dot\gamma_*(p),
\end{equation}
with
$\gamma_*(p)=\lim_{l\to\infty}
\gamma_l(p)$.
In $D=3$ we  obtain numerically $A_3\approx 1.075$.
In the critical long wavelength regime $y\gg 1$ one may
approximate all couplings by their initial values. 
The scaling function then approaches the constant value
\begin{equation}
\sigma_*(y)\approx\frac{2}{D}u^2_0 M^2_0  G^3_0(0).  
\label{eq:UpAsymptSigma}
\end{equation} 
Such a constant plateau is expected from the structure of the
truncation employed. In fact, for $k\gg k_c$ one expects
that the momentum dependence of the self-energy is that of lowest order
perturbation theory, see the discussion 
at the beginning of this subsection.
However, an effective ultra-violet cutoff is now provided by $k_c^{-1}$
which
regularizes the theory in place of the correlation length $\xi$ which is infinite
at criticality. While this is indeed the leading order correction
to the self-energy in an expansion in powers of the
bare interaction strength, the correct form of the self-energy
should further display a $\ln (k/k_c)$ dependence at large $k$ with
a pre-factor which is quadratic in the bare interaction \cite{Ledowski04,Hasselmann07}. 
The reason for the absence of
such a term in the present approximation is 
that our truncation for the four-point vertex
in (\ref{eq:G4initialtrunc}) does not take vertex corrections
into account.
In the macroscopically ordered regime  $k\xi\ll 1$  we find
that $ \Sigma ( \bd{k} ) - \Sigma (0 )$ vanishes as $\bd{k}^2$,
as can be seen in figure \ref{fig:SigmaLog}.

\begin{figure} \centering 
\includegraphics[height=5 cm]{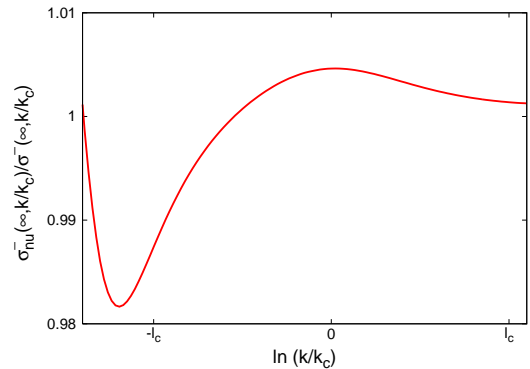}
\caption{ 
Comparison of two different approximations of the self-energy at criticality, as
discussed in the
text: self-consistent numerical solution ${\sigma}^-_{\rm nu} (\infty,k/k_c)$ 
of (\ref{eq:flowM}--\ref{eq:flowu})
and  solution $\sigma^{-} (\infty,k/k_c)$ based on the
substitution (\ref{eq:substit}) on the right-hand sides of these flow equations.
}
\label{diffplot}
\end{figure}

For the present model,
it is of course possible to calculate the 
solution of the
coupled integro-differential 
equations (\ref{eq:flowM}--\ref{eq:flowu}) exactly without
any further approximations. One obtains completely self-consistent
solutions for the flow of the self-energy $\Sigma_\Lambda(\bd{k})$ 
and the order parameter if the truncation of 
the momentum dependence
of ${\Sigma}_\Lambda$ and  ${G}_\Lambda$  on the right-hand
sides of (\ref{eq:flowM}--\ref{eq:flowu}) 
using the substitution (\ref{eq:substit}) is omitted.
A comparison of the completely self-consistent solution
${\sigma}_{\rm nu}^- (\infty,y)$
for the scaling function with the scaling function
$\sigma^{-}(\infty,y)$ obtained within the FRG enhanced perturbation
theory,
i.~e. with the help of the
substitution (\ref{eq:substit}) on the right-hand sides
of (\ref{eq:flowM}--\ref{eq:flowu}),
is shown in  figure \ref{diffplot}. Obviously, the  relative 
error due to the substitution  (\ref{eq:substit}) is
remarkably small,  
so that we conclude that our substitution (\ref{eq:substit}) is 
quite accurate.

Of course, one could easily improve  the approximations presented here.
A straightforward extension would be to truncate 
the effective potential $U_{\rm eff}$ in (\ref{eq:EffInt})
at some higher order $n>2$,
\begin{equation}
U_{\rm eff}[\varphi^2]\approx \sum_{m=2}^n  
\frac{u_\Lambda^{(m)}}{(2m)!} \big[\varphi^2-M_\Lambda^2\big]^m.
\end{equation}
To arrive at the flow equations
of the parameters $u_\Lambda^{(2)}, \dots, u_\Lambda^{(n)}$ one
would need to take into account the flow equation of the lowest $n$ vertices
of a field expansion. We have done so up to $n=5$; a significant
improvement of the critical exponents $\eta$ and $\nu$  
in the Ising model would however require approximately $n=10$,
as is known from previous investigations of the derivative
expansion~\cite{Canet03}. Note that our FRG enhanced perturbation theory
can also be adopted if
arbitrarily higher orders in $n$ are included. This is
because
within our approach
the flow of the coupling constants  which parameterize the local
potential depends only on the lowest order momentum expansion of
the self-energy and can
be calculated exactly as in the usual derivative 
expansion~\cite{Canet03}. Once the flow of the 
local potential is known,
the higher order momentum dependence of the
self-energy can be determined. Of course, not all models have a structure as
simple as the one discussed here which permits to include
arbitrarily high orders of the local potential. One may wonder
whether our approach is also useful to describe more complicated
systems.
It is certainly expected to be useful in non-critical interacting systems,
as e.~g. interacting bosons in two or three dimensions, where a low order
truncation of the effective action should suffice \cite{Sinner07}. 
An accurate 
description of the momentum dependent self-energy of more 
complicated and possibly critical systems, 
such as e.g. frustrated spin models \cite{Delamotte04}, 
is a very challenging
task which we have not yet attempted. While including all orders
of the local potentials would then be prohibitive, a low order
truncation might yet be qualitatively correct.


\section{Summary and conclusions}
\label{sec:CONC}

Let us summarize the three main results of this work:

\begin{enumerate}

\item
We have  demonstrated 
how
symmetry breaking can be included 
in the exact hierarchy of FRG flow equations for the irreducible vertices 
within the framework of the field expansion,
using the approach from \cite{Schuetz06}.
The basic idea is to require that the vertex with one external leg vanishes
identically for all values of the flow parameter, which yields an additional flow
equation for the order parameter.
Our method differs from that
employed in \cite{Salmhofer04,Honerkamp05,Gersch05}, 
which do not explicitly include a
flow equation for the order parameter.

\item Guided by the LPA, we have proposed a simple truncation of
the FRG flow equation for the momentum dependent self-energy $\Sigma ( \bd{k} )$, 
which yields a reasonable interpolation between the perturbative regime for large
momenta and the critical regime for $\bd{k} \rightarrow 0$.
We have also pointed out that  a sharp $\Theta$-function
cutoff (which is still very popular in the condensed matter community)
is not suitable to analyze the broken symmetry phase,
because it generates a non-analytic term proportional to $|\bd{k} |$ 
in the two-point vertex
for momentum scales below the
RG cutoff $\Lambda$

\item Using the above truncation, we 
have calculated the  two-parameter scaling
function $\sigma^{-} ( k \xi , k /k_c)$ describing the scaling of the self-energy
in the broken symmetry phase slightly below the critical temperature. 
 Similar to the case $T>T_c$  discussed in \cite{Hasselmann07},
the scaling function depends on two parameters,
involving the correlation length 
$\xi$ and the Ginzburg scale $k_c$. The latter remains finite at the critical
point and  measures  the size of the critical region.

\end{enumerate}

It is straightforward to generalize the
method described here to study spontaneous symmetry
breaking in
quantum mechanical many-body systems.
For example, our approach can be used to obtain fluctuation corrections
to the BCS gap equation in the attractive Fermi gas \cite{Lerch07}.
Note that beyond the BCS approximation the
superconducting order parameter should be distinguished from
the off-diagonal self-energy associated with
the single-particle Green function,
so that it is important to introduce
both quantities into the FRG as independent parameters.
We are currently using  our  method to study
the interacting Bose gas~\cite{Sinner07}, where
a truncation similar to the one discussed here  yields 
corrections to the Bogoliubov mean-field approximation for the diagonal and
off-diagonal self-energies which are consistent with the 
Hugenholtz-Pines theorem \cite{Nozieres90}.

\section*{ACKNOWLEDGMENTS}

We thank J. M. Pawlowski for useful discussions and acknowledge 
the collaboration with F. Sch\"{u}tz
at initial stages of this work.

\appendix
 
\setcounter{section}{1}

\section*{Appendix: Analytical form of $\dot\gamma_l(\bd{q})$  in three dimensions}
\label{appendix}

Using the  Litim cutoff \cite{Litim01}
given in (\ref{eq:RLitim})
we can rewrite the function
$\dot\gamma_l(\bd{q})$ defined in
(\ref{eq:Idef}) as 
\begin{eqnarray}
\dot\gamma_l(\bd{q})&=&  u^2_l  M^2_l  G^3_l(0)
\int_{{\bd{q}}^\prime} \; \dot{R}_l(\bd{q}^\prime) 
 \Theta\big(|\bd{q}^\prime+\bd{q}|^2-1\big) 
\frac{1-|\bd{q}^\prime+\bd{q}|^2} {|\bd{q}^\prime+\bd{q}|^2+ \rho_l}, 
\label{eq:GammaPoint}
\end{eqnarray}
where $\dot{R}(\bd{q})$ is defined in (\ref{eq:RPoint2}).
In $D=3$ the integration in (\ref{eq:GammaPoint})  can be performed 
exactly. We find 
\begin{equation}\label{eq:ExactGammaPoint}
\dot\gamma_l(\bd{q})=\Theta (2-q) \dot\gamma^{<}_{l}(q)+\Theta (q - 2) 
\dot\gamma^{>}_{l}(q),
\end{equation}
where the functions $\dot\gamma^{<}_{l}(q)$ and $\dot\gamma^{>}_{l}(q)$ 
are defined by 
\begin{eqnarray}\label{eq:GPpglt}
\fl \dot \gamma^{<}_l(q)= M^2_l  u^2_l  G^3_l(0)\bigg\{ A_l(q)+B_l(q) 
\ln\left[\frac{ \rho_l+(1+q)^2}{ \rho_l  + 1 }\right]
 \nonumber 
\\
-C_l(q)\left[\arctan\left(\frac{1}{\sqrt{ \rho_l}}\right)
-\arctan\left(\frac{1+q}{\sqrt{ \rho_l}}\right)\right]
\bigg\}, \\
\fl
\label{eq:GPpggt}
\dot \gamma^{>}_l(q)= M^2_l  u^2_l  G^3_l(0)\bigg\{ D_l(q)+E_l(q) 
\ln\left[\frac{ \rho_l+(1+q)^2}{ \rho_l+(1-q)^2}\right]
 \nonumber 
\\
+C_l(q)\left[\arctan\left(\frac{1-q}{\sqrt{ \rho_l}}\right)+\arctan\left(\frac{1+q}{\sqrt{ \rho_l}}\right)\right]
\bigg\},
\end{eqnarray}
with
\begin{eqnarray}\nonumber
\fl
A_l(q)=\frac{1}{480}\bigg\{
60\big[1+ \rho_l\big]\big[4-\eta_l(1+ \rho_l)\big]-30q \big[4-\eta_l+4  \rho_l(3-2\eta_l)
-7\eta_l  \rho^2_l\big]\nonumber \\ 
+20\eta_l 
q^2\big[5+9 \rho_l\big]-5q^3\big[4+\eta_l(17+25  \rho_l)\big]-2 \eta_l q^5\bigg\},
\\ 
\fl
B_l(q)=\frac{1+ \rho_l}{16 q}\bigg\{4\big[q^2- \rho_l-1\big]+\eta_l \big[ \rho^2_l+2  \rho_l (1-3 q^2)+(q^2-1)^2\big] \bigg\},
\\
\fl
C_l(q)=\frac{\sqrt{ \rho_l}}{2} \big(1+ \rho_l\big) \big[2-\eta_l (1+ \rho_l-q^2)\big],
\\
\fl
D_l(q)= \frac{\eta_l}{5} - \frac{1}{12} \big[\eta_l(5+ \rho_l)-8\big] -
( 1+ \rho_l) \bigl[ 1-\frac{\eta_l}{4} ( 2+3 \rho_l-q^2 ) \big],
\\
\fl
E_l(q)=\frac{1+ \rho_l}{16 q}\bigg\{ 
\eta_l-4+ \rho_l \big[\eta_l(2+ \rho_l)-4\big]-2q^2 \big[3\eta_l \rho_l+\eta_l-2 \big]+\eta_l q^4
\bigg\}.
\end{eqnarray}



\begin{thebibliography}{99}
\bibitem{footnote1}
As discussed  in \cite{Schuetz06},
it  is convenient to choose the free propagator such that
the product of the  inverse free propagator and the order parameter field
vanishes, $[G^{0}_{\Lambda} ( \bd{k} ) ]^{-1}  \varphi^0_{\bd{k}} =0$.
The factor of $1 - \delta_{ \bd{k} , 0 }$
in (\ref{eq:Rdef})
is introduced to satisfy this condition.
Without this factor
the right-hand side of (\ref{eq:orderflow})
would contain an additional term
$\partial_{\Lambda} [ R_{\Lambda} (0) M_{\Lambda} ] $.
Because the RG eliminates only modes with ${\bd{k}} \neq 0$ the projector
$1 - \delta_{ \bd{k} , 0 }$ does not affect the RG flow at any finite ${\bd{k}}$, while
the fluctuation of the zero mode $\Delta \varphi_{\bd{k} =0 }$ can be ignored
in the thermodynamic limit \cite{Schuetz06}.
\bibitem{footnote2}
Our factor $Z_{l}$ is the inverse of the corresponding
quantity $Z_k$ introduced in \cite{Berges02}.
\end{thebibliography}
\end{document}